\documentclass[12pt]{article}
\usepackage{multicol}
\usepackage{amssymb}
\usepackage{amsfonts,amssymb,amsmath,amsthm}
\usepackage{epsfig}
\usepackage{slashed}
\usepackage{graphicx}
\usepackage{float}
\usepackage{subfig}
\usepackage{mathrsfs}
\usepackage{bbm}
\usepackage{cite}
\usepackage{enumerate}
\usepackage{slashbox}
\usepackage{color}
\usepackage[colorlinks,linkcolor=red,anchorcolor=red,citecolor=green]{hyperref}
\makeatletter 

\makeatletter

\newcommand{\Rmnum}[1]{\expandafter\@slowromancap\romannumeral #1@}
\makeatother

\begin{document}
\renewcommand{\thefootnote}{\fnsymbol{footnote}}
\begin{titlepage}

\vspace{10mm}
\begin{center}
{\Large\bf Interaction potential and thermo-correction to the equation of state for thermally stable Schwarzschild Anti-de Sitter black holes}
\vspace{10mm}

{{\large Yan-Gang Miao${}^{}$\footnote{\em E-mail: miaoyg@nankai.edu.cn}
and Zhen-Ming Xu}${}^{}$\footnote{\em E-mail: xuzhenm@mail.nankai.edu.cn}

\vspace{3mm}
${}^{}${\normalsize \em School of Physics, Nankai University, Tianjin 300071, China}
}
\end{center}

\vspace{5mm}
\centerline{{\bf{Abstract}}}
\vspace{6mm}
The microscopic structure of black holes remains a challenging subject. In this paper, based on the well-accepted fact that black holes can be mapped to thermodynamic systems, we make a preliminary exploration of the microscopic structure of the thermodynamically stable Schwarzschild anti-de-Sitter (SAdS) black hole. In accordance with the number density and thermodynamic scalar curvature, we give the interaction potential among the molecules of thermodynamically stable SAdS black holes and analyze its effectiveness. Moreover, we derive the thermo-correction to the equation of state for such black holes that arises from interactions among black-hole molecules using virial coefficients.

\vspace{5mm}
\noindent
{\bf PACS Number(s)}: 04.70.Dy, 04.70.-s, 05.70.Ce

\vspace{5mm}

\end{titlepage}

\newpage
\renewcommand{\thefootnote}{\arabic{footnote}}
\setcounter{footnote}{0}
\setcounter{page}{2}
\pagenumbering{arabic}
\vspace{1cm}

\section{Introduction}
Black holes, due to their mysterious and elegant peculiarities, play a  unique role in connecting quantum mechanics and general relativity. In experimental observations,
human beings detected the ``sound'' of black holes for the first time using LIGO and VIRGO~\cite{GW}. This major breakthrough sheds light on the mystery of black holes. In theoretical research, the emergence of black-hole thermodynamics~\cite{SH,JDB,JMB,SHP,TP}, especially the issue on the extended phase space~\cite{CEJM,DSJT,BPD,KM,KM1}, injects a new vitality into the investigation of the properties of black holes. Black holes are gradually becoming the junction of quantum mechanics, general relativity, and statistical thermodynamics.

Based on the well-accepted fact that a black hole can be mapped to a thermodynamic system, one may conjecture that the black hole should have a microscopic structure even if its constituents are unknown. The exploration of this microscopic structure has progressed. In the early stage, string theory~\cite{st1,st2,st3,st4} and fuzzball theory~\cite{ft} were the most favorable candidates, 
wherein the relevant calculations depend either on supersymmetric and extremal configurations or on other speculations.
More recently, the concepts of black-hole molecules and their relevant number densities have been proposed~\cite{GR6}, allowing the small-large black hole phase transition of charged anti-de Sitter (AdS) black holes to be explained from the microscopic viewpoint. In particular, the concepts provide a new perspective for us to study the microstructure of black holes in the present work.
In addition, it has been found~\cite{BMS} that black-hole entropy can originate from the BMS symmetry at the horizon, which offers a new geometrical perspective. In particular, the microscopic mechanism of black-hole entropy can be explored from the viewpoint of thermodynamics. The spacetime atom approach~\cite{STA1,STA2,STA3} gives a possible microscopic description of gravity through a holographic equipartition law, whereas the Ruppeiner thermodynamic geometry~\cite{GR1,GR2,GR3,ALT,GR4,GR5,GR7,GR8,GR88,GR9,GR10,GR11,GR12} deals with the macroscopic properties of black holes as that of thermodynamic systems by extrapolation from the  black-hole-molecule hypothesis~\cite{GR6}.
In this study, we shall use the concepts of black-hole molecules and relevant number densities to analyze black-hole microstructure phenomenologically. The Ruppeiner thermodynamic geometry is closely related to the theory of fluctuations of equilibrium thermodynamics, where the Hessian is assumed~\cite{GR1,GR2,GR3,ALT} to be positive definite. From the thermodynamic scalar curvature, which is similar to that of general relativity, one can acquire~\cite{GR2} some qualitative information about the character of the internal molecular interaction for a thermodynamic system. That is, a positive (negative) thermodynamic scalar curvature implies a repulsive (attractive) interaction, whereas a vanishing thermodynamic scalar curvature implies no interaction.

Although there have been various attempts to explore the microscopic structure of SAdS black holes, the explicit construction of the microstates is still absent. It can be noted that the well-known Hawking-Page transition~\cite{SHP} exists in these types of black holes, and the phenomenon can be interpreted as the gravitational dual of the quantum chromodynamic confinement/deconfinement transition~\cite{EW1,EW2}. In this study, we shall attempt, from a phenomenological point of view, to give a possible picture of microscopic structure of SAdS black holes. In our previous works~\cite{GR10,GR11}, we took advantage of the Ruppeiner thermodynamic geometry to study the microscopic structures of a hairy black hole of Einstein's theory with a scalar field conformally coupled to higher-order Euler densities in high dimensions and to investigate the microscopic intermolecular potential for charged AdS black holes. Based on the concepts of black-hole molecules and black-hole intermolecular potentials, we shall intensively study the microscopic structure of SAdS black holes. First, we present the interaction potential among molecules of thermodynamically stable SAdS black holes and analyze its effectiveness in terms of the number density and the thermodynamic scalar curvature. Then, we derive the thermo-correction to the equation of state arising from interactions among black-hole molecules, which can be performed by utilizing the second virial coefficient.

This paper is organized as follows. In section \ref{sec2}, we briefly review some basic thermodynamic properties along with the Maxwell equal-area law for SAdS black holes in four dimensions. In section \ref{sec3}, we give the intermolecular interaction potential among SAdS black-hole molecules and analyze its feasibility in terms of the thermodynamic scalar curvature and the number density. We then derive the thermo-correction to the equation of state for the thermodynamically stable SAdS black hole in section \ref{sec4}. Finally, we draw our conclusion in section \ref{sec5}. Throughout this paper, we adopt the units $\hbar=c=G=1$.

\section{Thermodynamic properties}\label{sec2}
We proceed to review some basic thermodynamic properties of the SAdS black hole in four dimensions~\cite{BPD,ES1,BC}. The square of the line element is given by
\begin{eqnarray}
\text{d}s^2=-\left(1-\frac{2M}{r}+\frac{r^2}{l^2}\right)\text{d}t^2+\left(1-\frac{2M}{r}+\frac{r^2}{l^2}\right)^{-1}\text{d}r^2+
r^2(\text{d}\theta^2+\sin^2\theta\text{d}\phi^2),
\end{eqnarray}
where $M$ is the black-hole mass and $l$ represents the curvature radius of the AdS spacetime, which is related to the cosmological constant as $\Lambda=-3/l^2$.

The basic thermodynamic properties of SAdS black holes take the following forms in terms of the horizon radius, $r_h$, which is determined by the zero point of the $g^{-1}_{rr}$ component of the metric
\begin{eqnarray}
\text{Enthalpy}&:&M=\frac{r_h}{2}\left(1+\frac{r_h^2}{l^2}\right),\\
\text{Hawking temperature}&:&k_{_B}T_h=\frac{1}{4\pi r_h}\left(1+\frac{3r_h^2}{l^2}\right),\label{temp}\\
\text{Bekenstein-Hawking entropy}&:&\frac{S_{bh}}{k_{_B}}=\pi r_h^2,\label{entr}\\
\text{Pressure}&:&P=\frac{3}{8\pi l^2},\label{pres}\\
\text{(Thermo)Volume}&:&V=\frac43\pi r_h^3,\\
\text{Heat capacity}&:&C_{_P}=-2\pi r_h^2\left(1+\frac{3r_h^2}{l^2}\right)\left(1-\frac{3r_h^2}{l^2}\right)^{-1},\label{cap}
\end{eqnarray}
where $k_{_B}$ appearing in eqs.~(\ref{temp}) and~(\ref{entr}) is the Boltzmann constant. One can see that the Hawking temperature eq.~(\ref{temp}) has a minimum value of $k_{_B}T_{\text{min}}=\sqrt{3}/(2\pi l)$ at the horizon radius, $r_h=r_0\equiv l/\sqrt{3}$. For the heat capacity at constant pressure eq.~(\ref{cap}), there exist three cases: (i) when $r_h>r_0$, one can give a positive heat capacity,  $C_{_P}>0$; (ii) when $r_h<r_0$, one can deduce a negative heat capacity, $C_{_P}<0$, and (iii) when $r_h=r_0$, one can find that $C_{_P}$ is divergent. To remove the negative heat capacity, one can use the Maxwell equal-area law~\cite{ES1,BC,ESAS}. The reason for this is that ``the negative heat capacity at constant pressure describes unstable black holes corresponding to local maxima of the Gibbs free energy, which cannot stay in thermal equilibrium with the surrounding heat bath,'' as reported in ref.~\cite{ES1}. Hence, the thermodynamically stable black hole phase can be labeled~\cite{ES1} as the following form in which the Maxwell equal-area law is constructed in the $(T_h,S_{bh})$ plane,
\begin{align}
r_h&=\hat{r}\equiv\frac{l}{2\sqrt{3}}(\sqrt{13}-1),\label{hors}\\
\frac{S_{bh}}{k_{_B}}&=\frac{S}{k_{_B}}\equiv\frac{\pi l^2}{12}(\sqrt{13}-1)^2, \label{entrs}
\end{align}
and correspondingly, the Hawking temperature reads as~\cite{ES1}
\begin{eqnarray}
k_{_B}T_h=k_{_B}T\equiv\frac{\sqrt{3}}{12\pi l}(2\sqrt{13}-1). \label{temps}
\end{eqnarray}

Analyses of number density and interaction potential will be made in the next section, which is based on eqs.~(\ref{hors}), (\ref{entrs}), and (\ref{temps}). Herein, we comment on the three equations as follows.
\begin{itemize}
  \item In the $(T_h,S_{bh})$ plane (see Figure \ref{tu}) the SAdS black hole behaves similar to a van der Waals fluid in the $(P,V)$ plane.  As mentioned above, the negative heat capacity at constant pressure just corresponds to the unstable part that has had no credible physical explanation until now and should be removed.

\begin{figure}[H]
\begin{center}
\includegraphics[width=80mm]{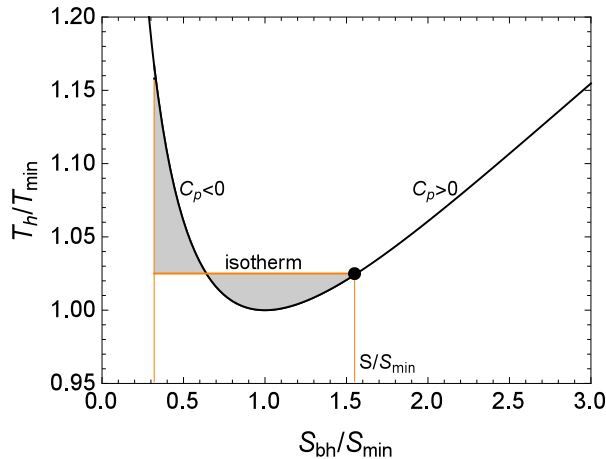}
\end{center}
\caption{The SAdS black hole in the $(T_h,S_{bh})$ plane and the construction of the Maxwell equal-area law. One can see that the temperature curve in the $C_p < 0$ region can be replaced by the isothermal process, which is made by requiring the areas of two shadow parts to be equal to one another.}
\label{tu}
\end{figure}

  \item The flattening of temperature through an isothermal process based on the construction of the Maxwell equal-area law shown in Figure \ref{tu} not only eliminates the unstable part, but also modifies the Hawking-Page phase transition, as reported in refs.~\cite{ES1,BC}.
  \item One has to apply the Maxwell equal-area law to choose a stable black hole when exploring its thermodynamic behaviors. In our case, eqs.~(\ref{hors}), (\ref{entrs}), and (\ref{temps}) describe a physically stable black hole obtained from the Maxwell equal-area law. This treatment is analogous to analysis of a small-large phase transition in charged AdS black holes~\cite{ESAS}.
\end{itemize}

\section{Number density and interaction potential}\label{sec3}
The number density is defined as $n\equiv 1/v$ in terms of the specific volume~\cite{KM}, $v\equiv 2r_h$, which is introduced~\cite{GR6,DCJ} to measure the microscopic degrees of freedom of black holes. Thus, the number density of a thermodynamically stable SAdS black hole takes the following form when we consider eqs.~(\ref{hors}) and~(\ref{temps}),
\begin{eqnarray}
n=\frac{\pi(\sqrt{13}-1)}{2\sqrt{13}-1}k_{_B}T. \label{num}
\end{eqnarray}

In contrast, the Ruppeiner geometry provides a powerful tool for investigating the microscopic properties of black holes completely from the thermodynamic point of view. The metric of this geometry can be written in the Weinhold-energy form~\cite{FW}
\begin{equation}
g_{\alpha\beta}=\frac{1}{k_{_B}T_h}\frac{\partial^2 M}{\partial X^{\alpha}\partial X^{\beta}}, \label{rg}
\end{equation}
where $X^{\alpha}$ takes $S_{bh}/k_{_B}$ and $P$, respectively, for the thermodynamic stable black hole. We can then obtain the thermodynamic scalar curvature,
\begin{equation}
R=-\frac{1}{8P(S_{bh}/k_{_B})^2+S_{bh}/k_{_B}}. \label{cur}
\end{equation}
Next, substituting eqs.~(\ref{pres}), (\ref{entrs}), and~(\ref{temps}) into eq.~(\ref{cur}), we can rewrite the thermodynamic scalar curvature as follows:
\begin{equation}
R=\mathcal{R}\equiv-2\pi\alpha (k_{_B}T)^2, \qquad \alpha=\left(\frac{\sqrt{13}+1}{2\sqrt{13}-1}\right)^3. \label{scal}
\end{equation}
One can clearly see that the thermodynamic scalar curvature is negative, i.e., $\mathcal{R}<0$, which coincides with the result for an ideal Bose gas~\cite{JM}. Therefore, by analogy with a thermodynamic system, we see that for the thermodynamically stable SAdS black hole, an attractive interaction dominates among black-hole molecules. Next, one may ask how this interaction is described quantitatively. We propose the simplest intermolecular interaction potential as a preliminary and tractable model, 
\begin{equation}
u(\mathfrak{r})=-4u_0 \left(\dfrac{\sigma}{\mathfrak{r}}\right)^6, \label{pote}
\end{equation}
where $\mathfrak{r}$ is center-of-mass separation between two molecules, $\sigma$ is the diameter of one molecule, and $u_0$ is a positive constant. Herein, we emphasize that the phrase ``intermolecular interaction'' is only a virtual description of black holes when they are treated as a thermodynamic system.
In addition, the minimum value of center-of-mass separation between molecules is $\mathfrak{r}_{min}=\sigma$, i.e., the molecules are hard in our model.

We now discuss in detail the validity of the above interaction potential for a thermodynamically stable SAdS black hole.
\begin{itemize}
  \item It is easy to obtain the interaction force of one molecule acting upon its neighbor from the molecular potential, eq.~(\ref{pote}),
\begin{eqnarray}
F(\mathfrak{r})=-\frac{24u_0}{\sigma}\left(\frac{\sigma}{\mathfrak{r}}\right)^7,
\end{eqnarray}
where $F(\mathfrak{r})$ is negative (meaning that the attractive interaction dominates). With increase of the center-of-mass separation $\mathfrak{r}$, the attractive interaction force $|F(\mathfrak{r})|$ decreases, whereas it increases with decrease of $\mathfrak{r}$. Note that the minimum separation is $\mathfrak{r}_{min}=\sigma$, i.e., $\mathfrak{r}\geq \sigma$.
  \item The number density is proportional to the Hawking temperature, see eq.~(\ref{num}). With the increase of Hawking temperature, the number density increases, implying that the center-of-mass separation between molecules is gradually decreasing.
  \item The thermodynamic scalar curvature is proportional to the square of the Hawking temperature, see eq.~(\ref{scal}). It is negative, which implies that an attractive interaction dominates among black-hole molecules. With increasing Hawking temperature, $|\mathcal{R}|$ also increases, indicating that the attractive interaction is becoming stronger.
  \item Because thermodynamically stable black holes are physically acceptable, our proposal, i.e., eq.~(\ref{pote}), only works well for such a thermodynamically stable SAdS black hole described by  eqs.~(\ref{hors}), (\ref{entrs}), and (\ref{temps}).
\end{itemize}

From the above mentioned analyses, we can summarize that the attractive interaction force $|F(\mathfrak{r})|$ increases with the increase of the Hawking temperature, whereas the thermodynamic scalar curvature $|\mathcal{R}|$ also increases. As a result, both descriptions of the interaction force $F(\mathfrak{r})$ agree with each other, where one is from the molecular potential, eq.~(\ref{pote}),  and the other is from the thermodynamic scalar curvature, $\mathcal{R}$, eq.~(\ref{scal}). Hence, the molecular potential, eq.~(\ref{pote}), is reasonable and acceptable for the exploration of the microscopic structure of the thermodynamically stable SAdS black hole from a thermodynamic point of view. Nonetheless, we emphasize that the molecular potential (eq.~(\ref{pote})) is in the simplest possible form, from which we can easily find  the thermo-correction to the equation of state. See the next section for details.

\section{Thermo-correction to the equation of state}\label{sec4}
In addition to the molecular potential, eq.~(\ref{pote}), which describes the microscopic structure of the thermodynamically stable SAdS black hole, 
we now derive the equation of state for the black hole dealt with as a thermodynamic system according to the statistical thermodynamic scheme. We utilize the virial equation of state~\cite{BOOK},
\begin{eqnarray}
P=\frac{k_{_B}T}{v}\left(1+\frac{B(T)}{v}+\frac{C(T)}{v^2}+\cdots\right), \label{ves}
\end{eqnarray}
where $v$ is the specific volume and the coefficients $B(T)$, $C(T)$, $\cdots$, are called the second, third, $\cdots$, virial coefficients. Using statistical mechanics, we can calculate these virial coefficients from the intermolecular potential functions. For the simplest case, the second virial coefficient can be extracted from the intermolecular potential as follows:
\begin{eqnarray}
B(T)=-\frac12 \int\left(e^{-u(\mathfrak{r})/(k_{_B} T)}-1\right)\text{d}\Omega,
\end{eqnarray}
where $\text{d}\Omega$ is volume element per molecule in the coordinate space. Using the molecular potential eq.~(\ref{pote}), we find the second virial coefficient in the spherical coordinate system $\{\mathfrak{r}, \theta, \varphi\}$,
\begin{align}
B(T) &=-\frac12 \int_{0}^{2\pi}\text{d}\varphi \int_{0}^{\pi}\sin\theta\text{d}\theta\int_{\sigma}^{\infty}\left(e^{-u(\mathfrak{r})/(k_{_B} T)}-1\right)\mathfrak{r}^2\text{d}\mathfrak{r} \nonumber\\
&=-2\pi\sigma^3\sum_{n=1}^{\infty}\frac{1}{n!(6n-3)}\left(\frac{4u_0}{k_{_B} T}\right)^n. \label{svir}
\end{align}
Herein, we comment on the above mentioned result.
\begin{itemize}
  \item For the thermodynamically stable SAdS black hole, we can derive the equation of state from eqs.~(\ref{temp}),~(\ref{pres}),  and~(\ref{temps}),
      \begin{eqnarray}
      P=\frac{k_{_B}T}{v}-\frac{1}{2\pi v^2}. \label{es}
      \end{eqnarray}
      For case $n=1$, substituting eq.~(\ref{svir}) into eq.~(\ref{ves}) and comparing the obtained result $P=k_{_B}T/v-8\pi u_0 \sigma^3/(3v^2)$ with eq.~(\ref{es}), we can obtain the relationship between the positive constant $u_0$ and the diameter of one molecule $\sigma$,
      \begin{eqnarray}
      u_0=\frac{3}{16\pi^2 \sigma^3}.
      \end{eqnarray}
  \item For cases $n\geq 2$, we can obtain the thermo-correction to the equation of state, eq.~(\ref{es}). This thermo-correction is induced by the intermolecular potential function eq.~(\ref{pote}). The corrected equation of state takes the following form,
      \begin{eqnarray}
      P=\frac{k_{_B}T}{v}-\frac{1}{2\pi v^2}\left[1+\sum_{n=2}^{\infty}\frac{1}{n!(2n-1)}\left(\frac{4u_0}{k_{_B} T}\right)^{n-1}\right]. \label{ces}
      \end{eqnarray}
  \item We can also see the highest order term in $1/v$ appearing in the equation of state eq.~(\ref{es}) is $1/v^2$. Hence, it is sufficient for us to consider only the correction of the second virial coefficient.
  \item According to eq.~(\ref{pote}), we can see that the maximum interaction potential energy is $4u_0$. In addition, the molecular average of thermal-kinetic energy is proportional to the factor $k_{_B}T$. For a thermodynamic system, we know that the average thermal-kinetic energy of the molecules is far greater than their potential energy, i.e., $k_{_B}T\gg 4u_0$. Turning to eq.~(\ref{ces}) again, we can say that the corrected terms are higher orders of the ratio $\frac{4u_0}{k_{_B}T}$, showing that the corrections are very small.
\end{itemize}

\section{Summary}\label{sec5}
From a strictly thermodynamic point of view, we have provided the intermolecular potential function (see eq.~(\ref{pote})) for the thermodynamically stable SAdS black hole.
Moreover, we discuss the validity of the interaction potential in terms of thermodynamic geometry. Our analyses show that the scheme of the molecular potential coincides with that of the thermodynamic geometry in the exploration of the microscopic structure of the thermodynamically stable SAdS black hole.
Furthermore, we derive the thermo-correction to the equation of state of the thermodynamically stable SAdS black hole by virtue of the second virial coefficient, see eq.~(\ref{ces}). Our investigation indicates that the thermally corrected terms are very small.
Incidentally, our treatment can be extended to other black-hole models, such as the Kerr black hole, the Kerr-Newman black hole, and the Guass-Bonnet AdS black hole, through calculation of higher virial coefficients.

\section*{Acknowledgments}
This work was supported in part by the National Natural Science Foundation of China under grant No.11675081. The authors would like to thank the anonymous referees for the helpful comments that improve this work greatly.

\end{document}